\documentclass[12pt]{article} \topmargin=0.5 cm  \textwidth 16.5
cm \textheight 23 cm \oddsidemargin 0pt \evensidemargin 0pt
\usepackage {graphicx}
\begin{document}

\title{\textbf{Effective potential analysis for 5D SU(2) gauge
       theories at finite temperature and radius.}}
\author{K. Farakos \footnote{E-mail: kfarakos@central.ntua.gr} and P. Pasipoularides \footnote{E-mail: paul@central.ntua.gr}
\\ \\  Department of Physics, National Technical University of
       Athens \\ Zografou Campus, 157 80 Athens, Greece}
\date{ }
       \maketitle

\begin{abstract}
We calculate the one loop effective potential for a 5D SU(2) gauge
field theory at finite temperature $T=1/\beta$ and radius $R=1/M$.
This calculation is performed, for the first time, in the case of
background fields with two constant components $A^{3}_{y}$
(directed towards the compact extra dimension with radius $R$) and
$A^{3}_{\tau}$ (directed towards the compact Euclidean time with
radius $\beta$). This model possesses two discrete symmetries
known as $Z_{M}(2)$ and $Z_{T}(2)$. The corresponding phase
diagram is presented in Ref. \cite{remark:2}. However the
arguments which lead to this diagram are mainly qualitative. We
present a detailed analysis, from our point of view, for this
phase diagram, and we support our arguments performing lattice
simulations for a simple phenomenological model with two scalar
fields interacting through the previously calculated potential.
\end{abstract}

\section{Introduction}
It has been noted long ago by G. 't Hooft in Ref. \cite{hooft:0}
that a pure 4D SU(N) gauge field theory at finite temperature $T$
develops a global symmetry which is known as $Z(N)$ symmetry. In
Refs. \cite{Weiss:01,Korthals:01}, the one loop effective
potential in the presence of a constant background gauge field
$A_{0}$ was calculated. This result, which is reliable only in the
weak coupling regime, implies a violation of the Z(N) symmetry.
This is interpreted as the phase transition to the deconfining
phase of an SU(N) gauge field theory, that is expected for high
temperatures.

In recent years, there has been an interest for models with extra
compact dimensions. The simplest way to extend the above model, is
to add an extra compact dimension $y$ with radius $R=1/M$. As a
consequence, the model develops an additional $Z(N)$ symmetry. To
distinguish the two $Z(N)$ symmetries of the model, we will call
$Z_{T}(N)$ the one that corresponds to the compact Euclidean time
and $Z_{M}(N)$ the one that corresponds to the extra compact
dimension (for details see the next section). Due to these
symmetries the model possesses four distinct phases.

A schematic phase diagram (see Fig. \ref{Iphase} below) has been
presented, for the first time, by C. P. Korthals Altes and M.
Laine in Ref. \cite{remark:2}. However the arguments, in Ref.
\cite{remark:2}, which lead to this phase diagram are mainly
qualitative. For this reason lattice simulations for a 5D and 4D
SU(2) gauge field theory at finite temperature and radius were
performed in Ref. \cite{lat:4}. The lattice results for $d=4$
confirm \footnote{The notation $d$ means that the model has d-2
noncompact and two compact dimensions.} the phase diagram of Fig.
\ref{Iphase}. Even in the case of $d=5$ where the theory is not
renormalizable, for fixed lattice spacing, the qualitative
features of the above mentioned phase diagram are evident in the
lattice results.

In this paper we calculate in detail the one loop effective
potential for a 5D SU(2) gauge field theory at finite temperature
$T$ and radius $R$. This calculation is performed in the case of
background fields with two constant components $A^{3}_{y}$ and
$A^{3}_{\tau}$.  This result generalizes a previous calculation,
in Ref. \cite{remark:2}, for one constant gauge field component
$A^{3}_{y}$ and $A^{3}_{\tau}=0$.

We will study whether it is possible to derive the qualitative
features of the phase diagram of Fig. \ref{Iphase} by using the
perturbative result for the effective potential. Unfortunately the
expected restoration of the $Z_{M}(N)$ symmetry \footnote{In this
work we analyze the N=2 case.} (when the system passes from region
(A) to (B) in Fig. \ref{Iphase}) for high temperature can not be
established just from the effective potential. For this reason we
construct a simple phenomenological model, that incorporates the
fluctuations of the scalar fields, by adding to the effective
potential kinetic terms. Numerical simulations on lattice for this
model give a phase diagram that exhibits the main features of the
expected phase diagram of Fig. \ref{Iphase}.

\section{The $Z_{T}(N)\times Z_{M}(N)$ symmetry.}

The object of study, in this section, is an SU(N) gauge field
theory, in a d-dimensional space-time at finite temperature, with
one extra compact dimension. This extra dimension will be noted by
$y$ and it varies from $0$ to $R=1/M$, where $M$ is the mass scale
of Kaluza-Klein modes. In the case of finite temperature $T$ we
have another compact dimension, the Euclidean time $\tau=it$,
which varies from 0 to $\beta=1/T$.

The partition function of this model is:
\begin{eqnarray}
 Z=\int_{b.c} {\cal D} A_{\mu} e^{- \frac{1}{2}\int_{0}^{R} dy \int_{0}^{\beta} d\tau \int d^{d-2}x
\; Tr\;(F_{\mu\nu})^{2}}
\end{eqnarray}
with the periodic boundary conditions
 \begin{eqnarray}
A_{\mu}(0,\tau,x)=A_{\mu}(R,\tau,x)\\
A_{\mu}(y,0,x)=A_{\mu}(y,\beta,x)
\end{eqnarray}
where $x=(x_{1},x_{2},...,x_{d-2})$. The field tensor is given by
the equation
$F_{\mu\nu}=\partial_{\mu}A_{\nu}-\partial_{\nu}A_{\mu}-i g_{d}
[A_{\mu},A_{\nu}]$ ($\nu,\mu=1,2...d$), where $g_{d}$ is the
d-dimensional coupling constant. In addition,
$A_{\mu}=A^{\alpha}_{\mu}T^{\alpha}$ ( $\alpha=1,2...N^{2}-1$),
and $T^{\alpha}$ satisfies the commutation relation
$[T^{\alpha},T^{b}]=i f^{abc}T^{c}$. Also note that for the
components $A_{d-1}$ and $A_{d}$ of the gauge field we will use
the notation $A_{\tau}$ and $A_{y}$.

We would like to emphasize that this action, if $d\geq 5$ (in this
work we will study the case of $d=5$), corresponds to a non
renormalizable field theory. However this theory is viewed as an
effective theory valid up to a finite cut-off $\Lambda$, and
describes the low energy behavior of a fundamental renormalizable
field theory, which may be a string theory. In this way all the
observables of this model are rendered finite. We note that an
observable like the one-loop effective potential, which is
computed in the next section, is finite and cut-off independent.
Also we note that the scale $\Lambda$ is assumed to be much larger
than the temperature $T$ and the mass scale $M$ (or $T<<\Lambda$
and $M<<\Lambda$). An extensive discussion on this topic is
presented by K. R. Dienes et al in Ref. \cite{dud:1}.

The action of this model is invariant under Gauge transformations
\begin{equation}
A'_{\mu}=UA_{\mu}U^{\dagger}+\frac{1}{ig_{d}}U\partial_{\mu}U^{\dagger}
\end{equation}
Note that the transformed gauge fields $A'_{\mu}$ should remain
periodic, otherwise we would have violation of the boundary
conditions (2) and (3) of the path integral. We see that these
conditions are satisfied if gauge transformations are also
periodic, namely $U(0,\tau,x)=U(R,\tau,x)$ and
$U(y,0,x)=U(y,\beta,x)$.

However the class of the gauge transformations that preserve
boundary conditions (2) and (3) is wider. In this class we have
also to include and the gauge transformations with the property:
\begin{eqnarray}
&&z_{1}U(0,\tau,x)=U(R,\tau,x) \\ &&z_{2}U(y,0,x)=U(y,\beta,x)
\end{eqnarray}
where $z_{1}$ and $z_{2}$ are elements of $Z(N)$. This means that
this model possesses an additional global discrete symmetry
$Z_{T}(N)\times Z_{M}(N)$, where $Z_{T}(N)$ corresponds to
Euclidean time $\tau$ and $Z_{M}(N)$ to the extra dimension $y$.
\begin{figure}[t]
\begin{center}
\includegraphics[scale=0.3,angle=0]{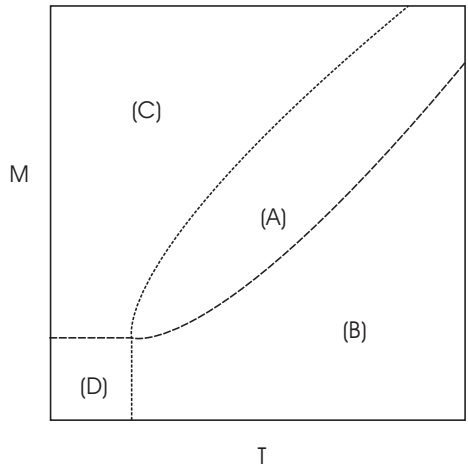}
\end{center}
\caption {The expected phase diagram for the 5D SU(N) at finite
temperature and radius. This phase diagram was proposed in Ref.
\cite{remark:2}, and it was confirmed by lattice simulations in
Ref. \cite{lat:4}. The regions \textbf{(A), (B), (C)} and
\textbf{(D)} in the figure are explained in the text below.}
\label{Iphase}
\end{figure}

Whether the symmetries $Z_{T}(N)$ and $Z_{M}(N)$ are violated or
not, depends on two order parameters $<P_{\tau}>$ and $<P_{y}>$
\footnote{We remind readers, that $<P_{\tau}>$ and $<P_{y}>$ are
the mean values of $P_{\tau}$ and $P_{y}$ in the corresponding
functional integral.}, where
\begin{eqnarray}
&&P_{\tau}(x,y)=\frac{1}{N} Tr{\cal P} e^{i
g_{d}\int_{0}^{\beta}d\tau A_{\tau}(y,\tau,x)}\\
&&P_{y}(x,\tau)=\frac{1}{N} Tr {\cal P} e^{i g_{d}\int_{0}^{R}dy
A_{y}(y,\tau,x)}
\end{eqnarray}
are the Polyakov loops in these directions.

Performing a Gauge transformation, with properties (5) and (6), we
see that the two order parameters are not invariant and transform
as: $<P_{\tau}>\rightarrow z^{\ast}_{1} <P_{\tau}>$ and
$<P_{y}>\rightarrow z^{\ast}_{2} <P_{y}>$. So depending on the
values of $T$ and $M$ we have four possible distinct phases:
\begin{description}
    \item[(A)] If $<P_{\tau}>\neq 0$ and $<P_{y}>\neq 0$ then both
the symmetries  $Z_{T}(N)$ and $Z_{M}(N)$ are violated, and then
we can use a low energy effective theory which is characterized as
3D SU(N)+adjoint matter.
    \item[(B)] If $<P_{\tau}>\neq 0$ and $<P_{y}>= 0$ then the symmetry
$Z_{T}(N)$ is violated but the symmetry $Z_{M}(N)$ is not
violated, and the theory is characterized as 4D SU(N)+adjoint
matter.
    \item[(C)] If $<P_{\tau}>=0$ and $<P_{y}>\neq 0$ then the
symmetry $Z_{T}(N)$ is not violated but the symmetry $Z_{M}(N)$ is
violated, and again the theory is characterized as 4D
SU(N)+adjoint matter.
    \item[(D)] If $<P_{\tau}>=0$ and $<P_{y}>=0$ then the symmetries
$Z_{T}(N)$ and $Z_{M}(N)$ are not violated, and so there is no low
energy effective theory description, then our theory is
characterized as 5D SU(N) theory.
\end{description}

As we see in Fig. \ref{Iphase} the M-T plane is separated into
four regions every one of which corresponds to one of the above
mentioned cases \textbf{(A)}, \textbf{(B)}, \textbf{(C)} and
\textbf{(D)}.

\section{One loop effective potential for 5D SU(2).}

In this section we will concentrate on the case of SU(2) for
$d=5$. We aim to compute the one loop effective potential in the
presence of a background field with two constant components
$A_{\tau}$ and $A_{y}$ which are directed toward the same
direction in the group space.

We split the gauge field into a classical background field
$B_{\mu}$ and a quantum field $\tilde{\alpha}_{\mu}$
$(\mu=1,2,3,4,5)$:
\begin{equation}
A_{\mu}=B_{\mu}+\tilde{\alpha}_{\mu}
\end{equation}
The background field $B_{\mu}$ is chosen to be zero in the case of
noncompact dimensions $x$ and constant for the compact dimensions
$\tau$ and $y$. \textit{We emphasize that in this work we study
only background fields with zero classical energy (or
$F_{\mu\nu}=0$)}. This happens only if we choose the gauge field
components $B_{\tau}$ and $B_{y}$ toward the same direction
(toward the generator $T_{3}$) in the group space. This choice is
also supported by the fact that it is a saddle point of the
constraint effective potential as we show in the appendix.

So the background field is chosen according to the following
equations:
\begin{equation}
B_{\tau}=\frac{2\pi T u}{g_{5}} T_{3}
\end{equation}
\begin{equation}
B_{y}=\frac{2\pi M q}{g_{5}} T_{3}
\end{equation}
where we have introduced the dimensionless scalar fields $q$ and
$u$.

We use a gauge-fixing condition of the form
$D^{\mu}\tilde{\alpha}_{\mu}^{\alpha}=0$, which is known as
background Feynman gauge, where
\begin{equation}
D^{a c}_{\mu}=\delta^{a c}\partial_{\mu}+g_{5}\varepsilon^{a b
c}B^{b}_{\mu}
\end{equation}
is the covariant derivative in the adjoint representation.

The lagrangian can be separated into three terms
$L_{FP}=\frac{1}{2}Tr\;(F_{\mu\nu})^{2}+L_{GF}+L_{GT}$, where
$L_{GF}=-\frac{1}{2}(D^{\mu}\tilde{\alpha}^{a}_{\mu})^{2}$ is the
gauge fixing term and $L_{GT}=\bar{\eta}^{a}((-D^{2})^{a
c}-D^{\mu} \varepsilon^{a b c}\tilde{\alpha}_{\mu}^{b})\eta^{c}$
is the ghost field term. If we keep only the quadratic terms in
the quantum fields we have:
\begin{equation}
L_{QT}=\frac{1}{2}\tilde{\alpha}^{a}_{\mu}\left[(-D^{2})^{a
c}\delta_{\mu\nu}\right]\tilde{\alpha}^{c}_{\nu}+\bar{\eta}^{a}(-D^{2})^{a
c}\eta^{c}
\end{equation}
Note that the linear term, in quantum fields, is identically zero
for the case of the background field of Eqs. (10) and (11), as it
is a solution of the equations of motion.

The effective potential is defined by the equation:
\begin{equation}
e^{-R\; \beta \;V \; V_{eff}(q,u)}=\int {\cal D}\bar{\eta}_{\mu}\;
{\cal D}\eta_{\mu} \; {\cal D}\tilde{\alpha}_{\mu}\;
e^{-\int_{0}^{R} dy \int_{0}^{\beta} d\tau \int  d^{d-2}x \;
L_{QT}}
\end{equation}
where $V$ is the space volume.

Integrating out the fluctuations $\tilde{\alpha}_{\mu}$ and the
ghost fields $\eta_{\mu}$, we obtain the following expression for
the effective potential:
\begin{equation}
V_{eff}(q,u)=\frac{1}{V}\;T\;M\;(\frac{d}{2}-1)\;Tr\ln(-D^{2})
\end{equation}
In order to perform the trace in the group space we write $-D^{2}$
in the following matrix form
\begin{equation}
 -D^{2}=\left(%
\begin{array}{ccc}
 - \partial_{\mu}^{2}+g_{5}^{2}(B_{\tau}^{3})^{2}+g_{5}^{2}(B_{y}^{3})^{2} & 2g_{5}B^{3}_{y} \partial_{y}+2g_{5}B^{3}_{\tau} \partial_{\tau}& 0 \\
 - 2g_{5}B^{3}_{y} \partial_{y}- 2g_{5}B^{3}_{\tau} \partial_{\tau} & -\partial_{\mu}^{2}+g_{5}^{2}(B_{\tau}^{3})^{2}+g_{5}^{2}(B_{y}^{3})^{2} &0  \\
  0 & 0 & -\partial_{\mu}^{2} \\
\end{array}%
\right)
\end{equation}
where we have used Eq. (12).

Taking into account Eqs. (10) and (11) we can write the
eigenvalues of the above matrix into the form
\begin{eqnarray}
&&\lambda_{1}=-\vec{\partial}^{2}-(\partial_{\tau}-i2\pi T
u)^{2}-(\partial_{y}-i2\pi M q)^{2} \nonumber \\
&&\lambda_{2}=-\vec{\partial}^{2}-(\partial_{\tau}+i2\pi T
u)^{2}-(\partial_{y}+i2\pi M q)^{2}\nonumber \\
&&\lambda_{3}=-\vec{\partial}^{2}-\partial_{\tau}^{2}-\partial_{y}^{2}\nonumber
\end{eqnarray}
For the trace in the functional space we will use a plane wave
basis $u(x,\tau,y)\sim e^{i\vec{p}\cdot\vec{x}} e^{i2 \pi T n
\tau}e^{i 2 \pi M m y} $ ($m,n=0,\pm 1,\pm 2,..$), then from Eq.
(15), if we renormalize by subtracting the effective potential
with no background field present (or $V_{eff}(q,u)\rightarrow
V_{eff}(q,u)-V_{eff}(0,0)$) , we obtain
\begin{eqnarray}
&&V_{eff}(q,u)= (d-2)TM\sum_{n,m=-\infty}^{+\infty}\int
\frac{d^{d-2}p}{(2\pi)^{d-2}}\ln\left(\frac{ p^{2}+(2\pi
M)^2(m+q)^{2}+(2\pi T)^{2}(n+u)^{2}}{p^{2}+(2\pi M)^{2}
m^{2}+(2\pi T)^{2} n^{2}}\right) \nonumber
\end{eqnarray}
From the integral representation
$\ln(a/b)=-\int_{0}^{+\infty}(ds/s)(e^{-as}-e^{-bs})$, we obtain:
\begin{eqnarray}
&&\ln\left(\frac{ p^{2}+(2\pi M)^2(m+q)^{2}+(2\pi
T)^{2}(n+u)^{2}}{p^{2}+(2\pi M)^{2} m^{2}+(2\pi T)^{2}
n^{2}}\right)\nonumber \\&&
=-\int_{0}^{+\infty}\frac{ds}{s}e^{-p^{2}s}[e^{-((2\pi
M)^2(m+q)^{2}+(2\pi T)^{2}(n+u)^{2})s}-e^{-((2\pi M)^{2}
m^{2}+(2\pi T)^{2} n^{2})s}]
\end{eqnarray}
If we perform first the integration over momentum we obtain:
\begin{equation}
V_{eff}(q,u)= (d-2)TM\frac{1}{(4\pi)^{(d-2)/2}}\int_{0}^{+\infty}
\frac{ds}{s^{d/2}}f(q,u,s)
\end{equation}
where
\begin{equation}
f(q,u,s)=-\sum_{n,m=-\infty}^{+\infty}[e^{-((2\pi
M)^2(m+q)^{2}+(2\pi T)^{2}(n+u)^{2})s}-e^{-((2\pi M)^{2}
m^{2}+(2\pi T)^{2} n^{2})s}]
\end{equation}

Using the Poisson formula
\begin{equation}
\sum_{n=-\infty}^{+\infty}F(n)=\sum_{r=-\infty}^{+\infty}[\int_{-\infty}^{+\infty}
dx e^{2 \pi i r x} F(x)]
\end{equation}
we obtain
\begin{equation}
f(q,u,s)=\sum_{r,l=-\infty}^{+\infty}\sqrt{\frac{1}{4 \pi
M^{2}s}}\sqrt{\frac{1}{4 \pi
T^{2}s}}e^{-r^{2}/(4M^{2}s)}e^{-l^2/(4 T^{2}s)}(1-e^{-2\pi i r q}
e^{-2 \pi i l u})
\end{equation}

Setting $s=1/(4 M^{2}\widehat{t})$
\begin{eqnarray}
V_{eff}(q,u)=(d-2)M^{d}\frac{1}{\pi^{d/2}}\sum_{r,l=-\infty}^{+\infty}\int_{0}^{+\infty}d\widehat{t}\;\widehat{t}^{d/2-1}e^{-(r^2+l^2/\rho^2)\widehat{t}}(1-e^{-2\pi
i r q} e^{-2 \pi i l u})
\end{eqnarray}
If we set $\widehat{t}=z/(r^2+l^2/\rho^2)$, and perform the
integration over z, we obtain
\begin{equation}
V_{eff}(q,u)=(d-2)M^{d}\frac{\Gamma(d/2)}{\pi^{d/2}}\sum_{r,l=-\infty}^{+\infty}\frac{1}{(r^2+l^2/\rho^2)^{d/2}}(1-e^{-2\pi
i r q} e^{-2 \pi i l u})
\end{equation}
where $\rho=T/M$, and we have used the equation:
$\int_{0}^{+\infty}dz \; z^{d/2-1}e^{-z}=\Gamma(d/2)$.

From Eq. (23) the one loop effective potential for the two scalar
fields can be put into the form:
\begin{equation}
V_{eff}(q,u)=V^{M}_{eff}(q)+V^{T}_{eff}(u)+V_{eff}^{int}(q,u)
\end{equation}
where
\begin{equation}
V^{M}_{eff}(q)=4 (d-2)
M^{d}\frac{\Gamma(d/2)}{\pi^{d/2}}\sum_{r=1}^{+\infty}
\sin^{2}(\pi r
q)[\frac{1}{r^{d}}+\sum_{l=1}^{+\infty}\frac{2}{(r^2+l^2/\rho^2)^{d/2}}]
\end{equation}

\begin{equation}
V^{T}_{eff}(u)=4 (d-2)
T^{d}\frac{\Gamma(d/2)}{\pi^{d/2}}\sum_{l=1}^{+\infty}
\sin^{2}(\pi l
u)[\frac{1}{l^{d}}+\sum_{r=1}^{+\infty}\frac{2}{(l^{2}+r^2
\rho^{2})^{d/2}}]
\end{equation}

\begin{equation}
V_{eff}^{int}(q,u)=-16 (d-2)M^{d} \frac{\Gamma(d/2)}{\pi^{d/2}}
\sum_{l=1}^{+\infty}\sum_{r=1}^{+\infty} \frac{\sin^{2}(\pi l
u)\sin^{2}(\pi r q)}{(r^2+l^2/\rho^{2})^{d/2}}
\end{equation}
Note that $V_{eff}(q,0)=V^{M}_{eff}(q)$ and
$V_{eff}(0,u)=V^{T}_{eff}(u)$.

\begin{figure}[h]
\begin{center}
\includegraphics[scale=1,angle=0]{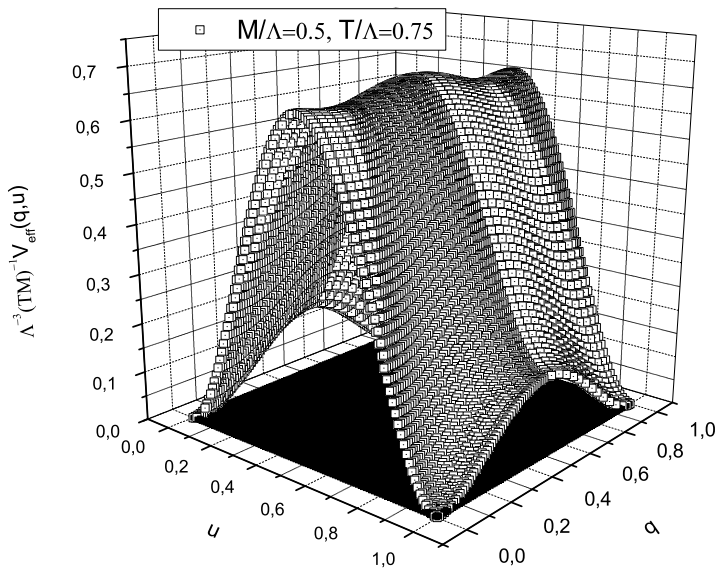}
\end{center}
\caption {$\Lambda^{-3}(TM)^{-1}V_{eff}(q,u)$ as a function of $q$
and $u$ for $M/\Lambda=0.5$ and $T/\Lambda=0.75$.} \label{pot3d}
\end{figure}

We observe that the potentials $V^{T}_{eff}(u)$ and
$V^{M}_{eff}(q)$  in Eqs. (25) and (26) are positive and the
potential $V_{eff}^{int}(q,u)$  in Eq. (27) is negative. One may
think that the effective potential $V_{eff}(q,u)$ in Eq. (24)
exhibits a local minimum for $q=1/2$ and $u=1/2$. However
numerical calculation, for several values of T and M in all
characteristic regions, shows that this is not the case. A typical
plot of the effective potential is shown in Fig. \ref{pot3d}.

An interesting feature of the effective potential is that it is
periodic in the dimensionless fields $q$ and $u$ with
corresponding periods equal to one. For this reason we have
plotted the $V_{eff}(q,u)$ only in the region $[0,1]\times[0,1]$.
Note that this periodicity of the effective potential is a
consequence of the $Z_{T}(2)\times Z_{M}(2)$ symmetry of the gauge
field theory.

We could not find an analytical expression for the effective
potential. However the effective potential $M^{-d}V^{M}_{eff}(q)$
(or $T^{-d}V^{T}_{eff}(u)$) can be approximated very well by a
function of the form $c \; q^{a}(1-q)^a$ where the parameters $c$
and $a$ are determined by a nonlinear fit procedure \footnote{We
have assumed that $q\epsilon[0,1]$. Strictly we should write
$M^{-5}V_{eff}(q)\approx c \; [q \; mod \;1]^{a}(1-[q\; mod \;
1])^a$. }. In Table \ref{prop} we present some values of $c$ and
$a$ for several $\rho$ for $d=5$. We see that as $\rho$ increases
the parameter $c$ increases linearly with $\rho$ and the parameter
$a$ tends to a constant value equal to $2$.

Now for the special case of $\rho=0$ we have:
\begin{equation}
M^{-5}V^{M}_{eff}(q)=\frac{9}{\pi^{2}}\sum_{r=1}^{+\infty}
\frac{\sin^{2}(\pi r q)}{r^{5}}= 18.87 \; q^{2.183}(1-q)^{2.183}
\end{equation}
For the case of $\rho \rightarrow +\infty$ we have shown,
performing accurate numerical computations, that:
\begin{equation}
\frac{1}{r^{5}}+\sum_{l=1}^{+\infty}\frac{2}{(r^2+l^2/\rho^2)^{5/2}}\sim
\frac{4}{3}\rho \frac{1}{r^{4}}
\end{equation}
Thus from Eqs. (29) and (25)
\begin{equation}
M^{-5}V^{M}_{eff}(q)\sim\frac{12}{\pi^{2}}\rho\sum_{r=1}^{+\infty}
\frac{\sin^{2}(\pi r q)}{r^{4}} \sim 2 \pi^{2} \rho \;
q^{2}(1-q)^{2}
\end{equation}
where we have used
$x^{2}(1-x)^{2}=\frac{6}{\pi^{4}}\sum_{r=1}^{+\infty}
\frac{\sin^{2}(\pi r x)}{r^{4}}$ (for this formula see for example
Ref. \cite{two:5}).

\begin{table}
  \centering
  \begin{tabular}{|c|c|c|c|}
    \hline
    $\rho$ & c & c/$\rho$ & $a$  \\
    \hline
    0    &   18.87   &-& 2.183 \\
    0.5 & 17.78 & 35.57 &2.110\\
    1   &21.11 & 21.11  & 2.015 \\
    2   &39.49 & 19.74 & 2.000 \\
    3 &59.22 & 19.74  & 2.000 \\
    4  & 78.96 & 19.74 & 2.000   \\
    \hline
  \end{tabular}
  \caption{A very good approximation for the effective potential $M^{-5}V^{M}_{eff}(q)$
  is given by a curve of the form $c \; q^{a}(1-q)^a$. The parameters $c$ and $a$ are
   determined with a nonlinear fit procedure for several values of $\rho$ and are presented in the above table.
We see that the parameter $c$ is proportional to $\rho$ for
$\rho>>1$. The relative errors for the parameters $a$ and $c$ are
of the order of a thousandth or smaller and are not presented in
this table.}\label{prop}
\end{table}

According to Fig. \ref{Iphase} we expect a restoration of
$Z_{M}(2)$ symmetry above a temperature $T_{c}$ ($T_{c}>M$) where
the system passes from region (A) to region (B). However the
perturbative results can not explain the restoration of the
$Z_{M}(2)$ symmetry for large temperatures, as the barrier that
separates the vacua $q=0$ and $q=1$ increases linearly with the
temperature (this is also emphasized in Ref. \cite{remark:2}).

\section{A 3-dimensional model with two scalar fields.}

In this section we will present an analysis, from our point of
view, for the $Z_{M}(2)$ symmetry restoration and more generally
for the phase diagram of Fig. \ref{Iphase}, taking into account
the result for the effective potential of Eq. (24).

We consider that the fields $u=g_{5}B^{3}_{\tau}/2 \pi T$ and
$q=g_{5}B^{3}_{y}/2 \pi M$, are not constant, as it was assumed,
but they are dependant on the spatial coordinates
$x=(x_{1},x_{2},x_{3})$. Now we can construct a new action by
adding to the one loop effective potential the kinetic terms which
are obtained by substituting Eqs. (10) and (11) in the original
action of the five dimensional gauge theory. Then we have:
\begin{equation}
S_{eff}[q,u]=\frac{2 \pi^{2}}{ g_{5}^{2} \rho}\int d^{3}x \;
(\partial q)^{2}+\frac{2 \pi^{2} \rho}{ g_{5}^{2}} \int d^{3}x\;
(\partial u)^{2}+\frac{1}{T M}\int d^{3}x \; V_{eff}(q,u)
\end{equation}
This simple model is viewed as a quantum field theory and the
expectation value of an observable quantity $\hat{O}(q,u)$ (an
example of an observable quantity is the Polyakov loop in Eq. (38)
below) is obtained by the path integral
\begin{equation}
\langle \hat{O}(q,u)\rangle =\frac{\int {\cal D}q {\cal D}u \;
\hat{O}(q,u)\; e^{-S_{eff}(q,u)}}{\int {\cal D}q {\cal D}u \;
e^{-S_{eff}(q,u)}}
\end{equation}

Note that this model is nonrenormalizable as the potential is
periodic and thus includes powers of $q$ and $u$ larger than $six$
\footnote{We remind readers, that a 3 dimensional scalar field
theory that includes powers of the scalar field up to four is
superenormalizable. If the powers of the scalar field are up to
six the theory is renormalizable else the theory is
nonrenormalizable, see for example Ref. \cite{ren:6}.}. However,
in this paper, our model is viewed as a low energy effective
theory that is valid up to a finite momentum cut-off $\Lambda$. In
this way all observable quantities are rendered finite. The
momentum cut-off $\Lambda$ could be identified with the momentum
cut-off of the original gauge theory.

\textit{Our purpose, for the introduction of this scalar model, is
to incorporate fluctuations for the scalar fields $q$ and $u$}. An
estimation of the intensity of fluctuations is given by the
inverse of the coefficients in the kinetic terms of Eq.(31). The
fluctuations for the field $u$ are controlled by the parameter
$g^{2}_{5}/\rho$ and for the field $q$ by $g^{2}_{5}\rho$ where
$\rho=T/M$. So, for example, when we increase the temperature $T$
keeping the mass scale $M$ fixed we increase the fluctuations for
$q$ and suppress the fluctuations for $u$.

Note that the model possesses four topologically nonequivalent
vacua: $(q=0,u=0)$, $(q=0,u=1)$, $(q=1,u=0)$, $(q=1,u=1)$. These
vacua are separated by potential barriers, as we see in  Fig.
\ref{pot3d}. According to the above model, the system can jump
from one vacuum to another only due to fluctuations of the
dimensionless scalar fields $q$ and $u$.

When our system is in region (A) (see Fig. \ref{Iphase}) $q$ and
$u$ are frozen to one of the four vacua of the model. As the
temperature T increases, and the mass scale M is kept fixed, the
barrier between the vacua $(q=0,u=0)$ and $(q=0,u=1)$ (or
$(q=1,u=0)$ and $(q=1,u=1)$ ) increases rapidly (like $T^{5}$ as
we obtain from Eq. (26)). In addition the fluctuations for the
field $u$ are getting weaker, thus they can not help the system to
jump from one vacuum to another and the field $u$ remains frozen
to $u=0$ or to $u=1$.

On the other hand, the barrier between the vacua $(q=0,u=0)$ and
$(q=1,u=0)$ (or equivalently between $(q=0,u=1)$ and $(q=1,u=1)$),
as we see from Tab. \ref{prop}, is proportional to $\rho=T/M$. In
this case, as it is remarked in the last paragraph, the
fluctuations of the field $q$ are getting stronger as the
temperature increases, and this can help the system to jump from
one vacuum to another and may have as a result the restoration of
the $Z_{M}(2)$ symmetry (then the system passes from region (A) to
region (B) in Fig. \ref{Iphase}).

\section{The lattice model}

Our aim is to perform numerical computation and to see if the
model can confirm the basic features of the phase diagram of Fig.
\ref{Iphase}. The only way to use this model for numerical
computations is to discretise the action of Eq. (31) in the
lattice.

We will denote the lattice points by $n=(n_{1},n_{2},n_{3})$ where
$n_{1},n_{2},n_{3}$ are integers. If the lattice spacing will be
denoted by $a$, then the corresponding physical points are
$x_{n}=(a n_{1},a n_{2},a n_{3})$. We remind readers that in this
model we assume fixed lattice spacing which is set equal to the
inverse value of the momentum cut-off $\Lambda$ of the five
dimensional gauge theory.

The lattice action reads:
\begin{eqnarray}
S_{eff}[q,u]=&\beta_{g}&\frac{\pi^{2}}{2}\sum_{n}\sum
_{\mu=1}^{3}\left(\frac{1}{\rho}(q(x_{n}+a e_{\mu})-q(x_{n}))^{2}+
\rho \; (u(x_{n}+a e_{\mu})-u(x_{n}))^{2} \right)\nonumber
\\&+&\sum_{n} \; \hat{V}_{eff}(q(x_{n}),u(x_{n}))
\end{eqnarray}
where $\beta_{g}=4 a/ g_{5}^{2}$, $e_{\mu}$ are the unit vectors,
and we choose to measure all the quantities in the action in units
of the lattice spacing.

The potential $\hat{V}_{eff}(q,u)$ is defined as
$\hat{V}_{eff}(q,u)=\frac{a^{3}}{T M}V_{eff}(q,u)$. According to
this definition we have:
\begin{eqnarray}
\hat{V}_{eff}(q,u)=\hat{V}^{M}_{eff}(q)+\hat{V}^{T}_{eff}(u)+\hat{V}_{eff}^{int}(q,u)
\end{eqnarray}
where
\begin{equation}
\hat{V}^{M}_{eff}(q)=4 (d-2)
m^{d-2}\rho\frac{\Gamma(d/2)}{\pi^{d/2}}\sum_{r=1}^{+\infty}
\sin^{2}(\pi r
q)[\frac{1}{r^{d}}+\sum_{l=1}^{+\infty}\frac{2}{(r^2+l^2/\rho^2)^{d/2}}]
\end{equation}

\begin{equation}
\hat{V}^{T}_{eff}(u)=4 (d-2)
t^{d-2}(1/\rho)\frac{\Gamma(d/2)}{\pi^{d/2}}\sum_{l=1}^{+\infty}
\sin^{2}(\pi l
u)[\frac{1}{l^{d}}+\sum_{r=1}^{+\infty}\frac{2}{(l^{2}+r^2
\rho^{2})^{d/2}}]
\end{equation}

\begin{equation}
\hat{V}_{eff}^{int}(q,u)=-16 (d-2)m^{d-2}\rho
\frac{\Gamma(d/2)}{\pi^{d/2}}
\sum_{l=1}^{+\infty}\sum_{r=1}^{+\infty} \frac{\sin^{2}(\pi l
u)\sin^{2}(\pi r q)}{(r^2+l^2/\rho^{2})^{d/2}}
\end{equation}
Note that we have set  $t=T a$ and $m=M a$.

The order parameters, according to which the symmetries $Z_{T}(2)$
and $Z_{M}(2)$ are violated or not, are defined by the average
values of the Polyakov loops (see Eqs. (7) and (8)) which, for the
lattice model we examine, are given by the equations:
\begin{equation}
P_{\tau}(x)=\cos(\pi u(x)), \quad P_{y}(x)=\cos(\pi q(x))
\end{equation}
The volume averages of the Polyakov loop are:
\begin{equation}
\bar{P}_{\tau}=\frac{1}{V_{L}}\sum_{\{n\}} P_{\tau}(x_{n}), \quad
\bar{P}_{y}=\frac{1}{V_{L}}\sum_{\{n\}} P_{y}(x_{n})
\end{equation}
and $V_{L}=V/\alpha^{3}$ is the lattice volume.

So with the lattice simulation we will measure the quantities:
\begin{equation}
|P_{\tau}|=\langle  |\bar{P}_{\tau}| \rangle, \quad
|P_{y}|=\langle |\bar{P}_{y}| \rangle
\end{equation}

The critical values of $t$ (or critical values of $m$) are specified
as the values which maximize the susceptibilities:
\begin{equation}
\chi(P_{\tau})=V_{L}\left[\langle
(\bar{P}_{\tau})^{2}\rangle-(\langle  |\bar{P}_{\tau}|
\rangle)^{2} \right], \quad \chi(P_{y})=V_{L}\left[\langle
(\bar{P}_{y})^{2}\rangle-(\langle  |\bar{P}_{y}| \rangle)^{2}
\right]
\end{equation}

\section{Numerical results}

\begin{figure}[h]
\begin{center}
\includegraphics[scale=1,angle=0]{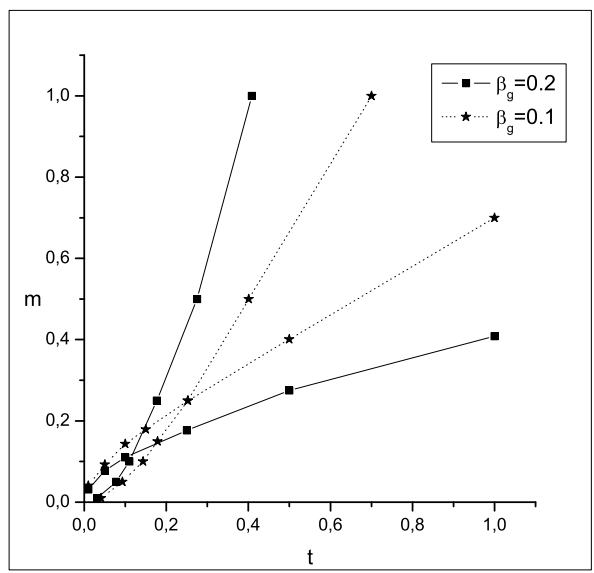}
\end{center}
\caption {The phase diagram of the model for fixed $\beta_{g}=0.2$
or $0.1$, for finite lattice volume $V_{L}=16^{3}.$} \label{num1}
\end{figure}
\begin{figure}[h]
\begin{center}
\includegraphics[scale=1,angle=0]{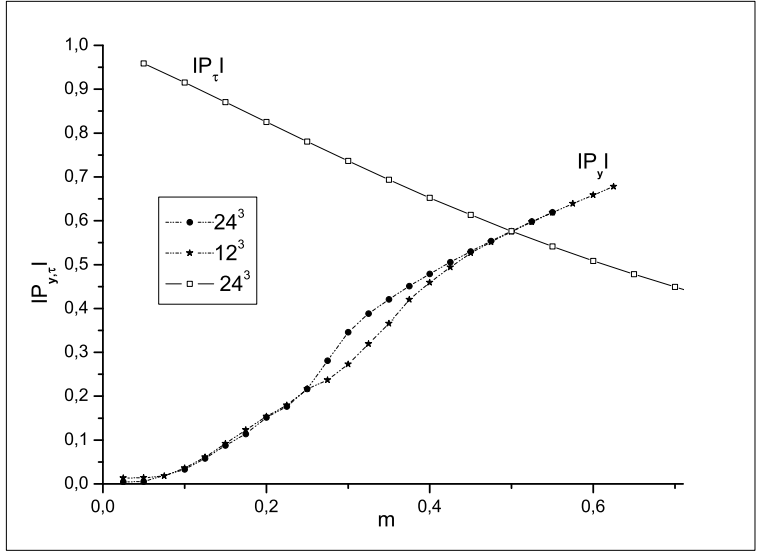}
\end{center}
\caption {$|P_{y}|$ and $|P_{\tau}|$ as a function of $m$ for
$\beta_{g}=0.2$, $t=0.5$ and $V_{L}=12^{3}, 24^{3}$. We see that
as the lattice volume increases the phase transition, which
corresponds to $|P_{y}|$ , is getting more sharp. The phase
transition for $|P_{\tau}|$ is not shown in this plot.}
\label{num2}
\end{figure}
\begin{figure}[h]
\begin{center}
\includegraphics[scale=1,angle=0]{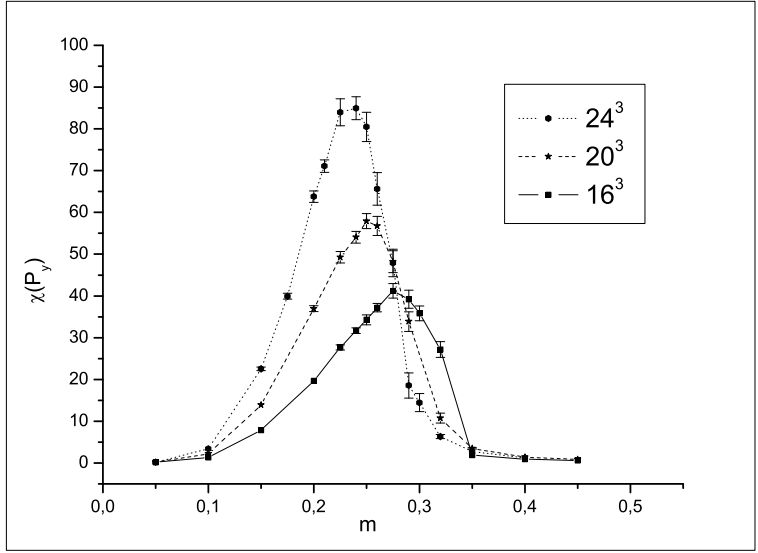}
\end{center}
\caption { $\chi(P_{y})$ as a function of $m$ for $\beta_{g}=0.2$,
$t=0.5$ and $V_{L}=16^{3},20^{3}$ and $24^{4}$. As the lattice
volume increases $m_{c}$ is moving slowly towards the left and,
for large lattice volume $V_{L}$, it seems to tend to an
asymptotic value.} \label{num3}
\end{figure}
\begin{figure}[h]
\begin{center}
\includegraphics[scale=1,angle=0]{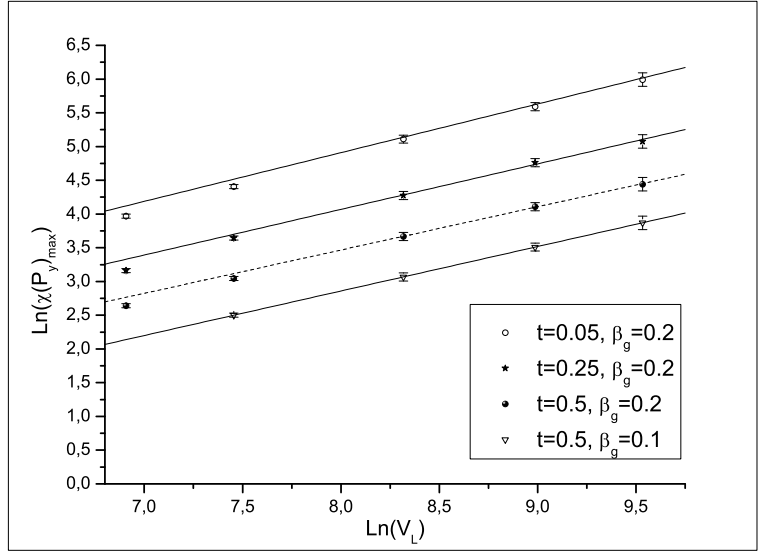}
\end{center}
\caption {$Ln(\chi(P_{y})_{max})$ as a function of $Ln(V_{L})$.
The discrete points corresponds to $V_{L}=10^{3}$, $12^{3}$,
$16^{3}$, $20^{3}$ and $24^{3}$. The continue line corresponds to
a best fit curve of the form $Ln(\chi_{max})=a+b\cdot Ln(V_{L})$
for the three bigger volumes.} \label{num4}
\end{figure}
\begin{figure}[h]
\begin{center}
\includegraphics[scale=1,angle=0]{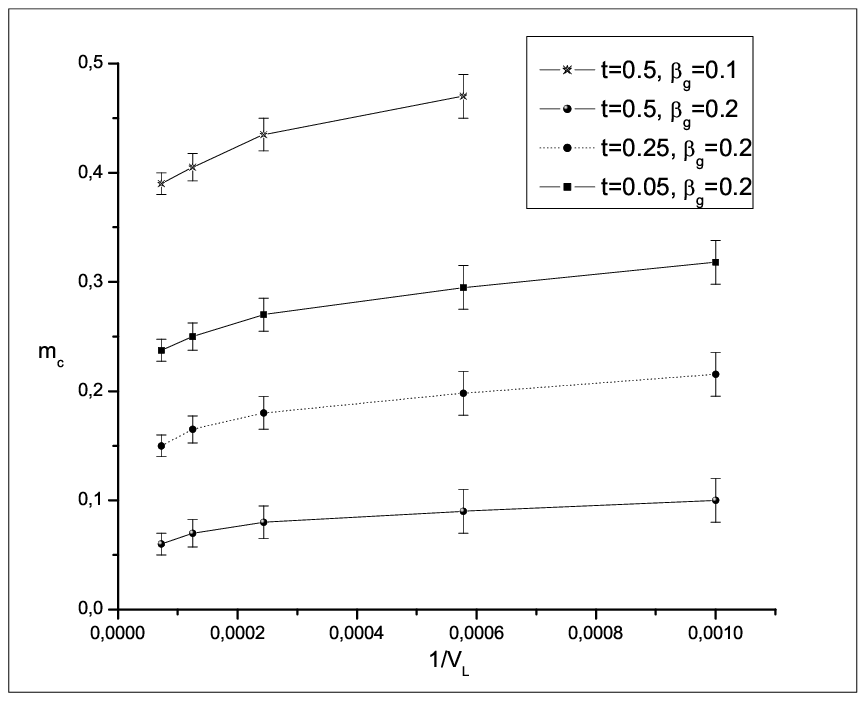}
\end{center}
\caption {$m_{c}$ for fixed $\beta_{g}=0.2$, $t=0.5, 0.25, 0.05$
and $\beta_{g}=0.1$, $t=0.5$ for five distinct values of lattice
volume $10^{3}$, $12^{3}$, $16^{3}$, $20^{3}$ and $24^{3}$. The
lines are there to guide the eye.} \label{num5}
\end{figure}

In this section we will study numerically the phase diagram $t-m$,
for fixed $\beta_{g}$. For this we have performed lattice
simulation for several lattice volumes $V_{L}=10^{3}, 12^{3},
16^{3}, 20^{3}$ and $24^{3}$.  Near the peaks of the
susceptibilities, for $V_{L}=24^{3}$, where the phase transition
happens, we have used samples of 250K measurements which are
separated by nine Metropolis iterations. The first 20K
measurements were ignored for thermalization. For the other
lattice volumes we have used samples with fewer measurements.

In Fig. \ref{num1} we have plotted the phase diagram for
$\beta_{g}=0.2$ and $0.1$ for finite lattice volume
$V_{L}=16^{3}$. We see that it exhibits the main features of Fig.
\ref{Iphase}, namely it separates the $t-m$ plane into the four
distinct regions that correspond to the four phases (A),(B),(C)
and (D) of the theory.

The order parameter $|P_{y}|$ as a function of $m$, for
$\beta_{g}=0.2$ and $t=0.5$ is shown in Fig. \ref{num2}. We
observe that the phase transition of the system remains continuous
even for large volumes $V_{L}=24^{3}$. Note that we have performed
computations for several other characteristic values of $t$ and
$\beta_{g}$ and the continuous behavior of the order parameter as
a function of $m$ is the same.

The corresponding susceptibilities are shown in Fig. \ref{num3}.
The location of the points of the phase transition and the
estimation of the relative errors have been done by computing the
susceptibility several times in the range where the peak is
expected. However we have not used a histogram method as it is not
applicable to the model we examine.

\textit{From the behavior of the peaks of the susceptibilities it
seems that we have a second order phase transition.} According to
the theory of finite size scaling (see for example Ref.
\cite{Monte:7}) we expect that the peak depends on the lattice
volume as $\chi(P_{y})_{max}= c V_{L}^{b}$ for large values of
$V_{L}$, where $b=\gamma/3\nu$ (for the definition of the critical
exponents $\gamma$ and $\nu$ see Ref. \cite{Monte:7}). This
behavior was confirmed numerically, for the model we examine, and
it is shown in Fig. \ref{num4}. The errors in the figure were
estimated by the Jackknife method. The critical exponent $b$ was
determined by a linear fit and it was found to be, for
$\beta_{g}=0.2$ , $b=0.70\pm 0.04$ for $t=0.5$, $b=0.68\pm 0.04$
for $t=0.25$ and $b=0.64\pm 0.04$ for $t=0.05$ using the three
bigger volumes. For $\beta_{g}=0.1$ and $t=0.5$ we found that
$b=0.66\pm 0.04$. These numerical values for $b$ ($0<b<1$)
indicate that we do not have a first order phase transition and
possibly the phase transition is of second order. Moreover, this
is a strong indication that these phase transitions belong to the
same universality class, as the numerical values of these critical
exponents are very close and lie in the range of errors. Finally,
we note, that these values for $b$ could be equal to the
corresponding critical exponent of the 3d ising model $b=0.657$.
This numerical value for the critical exponent $b$ was obtained in
Ref. \cite{peskin:13}.

We argue also that the qualitative features of the phase diagram,
of Fig. \ref{num1}, can not be just finite size effects. For this
we have plotted in Fig. \ref{num5} the critical value $m_{c}$ for
several values of $t=0.05,0.25$ and $0.5$, for $\beta_{g}=0.2$ and
$0.1$, as a function of lattice volume $V_{L}$. We see that there
are small displacements for $m_{c}$ but the arrangement of the
critical values does not change as the the lattice volume
increases. This indicates that, in the infinite lattice volume
limit, the qualitative features of the phase diagram are
preserved.

Note that we have not used the data points in Fig. \ref{num5} in
order to to determine the critical exponent, as a fitting of the
form $m_{c}=m_{\infty}+c'/V_{L}^{1/3 \nu}$ will give unreliable
results. The reason is that we have to determine three independent
parameters whose values are very sensitive and we have not enough
data points with a satisfactory accuracy.

\section{Conclusions}

We have computed the one loop effective potential for a 5D SU(2)
gauge field theory at finite temperature and radius in the case of
a background field with two constant components $A^{3}_{y}$ and
$A^{3}_{\tau}$.

However the effective potential, which is a perturbative result,
can not explain straightforwardly all the qualitative features of
the phase diagram of Fig. \ref{Iphase}. For this we constructed a
phenomenological model by adding to the one loop effective
potential kinetic terms (see Eq. (31)). We performed Monte Carlo
simulations with the lattice version of this model (see Eq. (33))
and we found a phase diagram, for fixed lattice spacing and
$\beta_{g}$, which exhibits all the qualitative features of Fig.
\ref{Iphase}.

Now the restoration of $Z_{M}(2)$ for large temperatures (or the
passing from region (A) to region (B) in Fig. \ref{Iphase}) can be
understood in the following simplistic way: even though the
barrier that separates the vacua of the dimensionless field $q$
increases linearly with the temperature $T$, the fluctuations of
$q$ are getting stronger (see section 4), and as it is confirmed
by the lattice model, it succeeds in restoring the $Z_{M}(2)$
symmetry. At the same time the fluctuations of $u$ are getting
more and more restricted so the field $u$ is frozen to one of its
vacuum states.

Finally we remark that the numerical results indicate second order
phase transitions, which is an interesting feature of the lattice
model of Eq. (33). However, questions like the continuous limit
(or nonperturbative renormalizability) are beyond the scope of
this paper.

\section{Acknowledgements}

We would like to thank P. Dimopoulos  and K. Anagnostopoulos for
reading and comments on the manuscript. We also thank G.
Tiktopoulos for valuable discussions, and C. P. Korthals Altes for
reading the manuscript and his comments on the constraint
effective potential. The work of P.P. was supported by the
"Pythagoras" project of the Greek Ministry of Education. The
numerical computations have been carried on the cluster of the
Physics Department of NTUA.

\section{Appendix: The constraint effective potential for two scalar fields}
Generalizing the definition of the constraint effective potential
$\widetilde{V}_{eff}$ in Ref. \cite{two:5} for the case of two
order parameters we have:
\begin{equation}
 e^{-V\beta R \widetilde{V}_{eff}(t_{4},t_{5})}=\frac{1}{Z}\int {\cal
 D}A_{\mu}^{\alpha}\delta(t_{4}-\bar{t}_{4})\delta(t_{5}-\bar{t}_{5})e^{-S[A_{\mu}^{\alpha}]}
\end{equation}
where
\begin{equation}
\bar{t}_{4}(A_{4}^{\alpha})=\frac{1}{2VR}\int_{0}^{R}dy \int
d^{3}x \; Tr P e^{ig_{5}\int_{0}^{\beta}A_{4}(x,\tau,y)d\tau}
\end{equation}
and
\begin{equation}
\bar{t}_{5}(A_{5}^{\alpha})=\frac{1}{2 V
\beta}\int_{0}^{\beta}d\tau \int d^{3}x \; Tr P
e^{ig_{5}\int_{0}^{R}A_{5}(x,\tau,y)dy}
\end{equation}
If we use the following representations for the delta functions
\begin{equation}
 \delta(t_{4}-\bar{t}_{4})=\int_{-\infty}^{+\infty}d\lambda_{4}e^{i\lambda_{4}(t_{4}-\bar{t}_{4})}
\end{equation}
and
\begin{equation}
 \delta(t_{5}-\bar{t}_{5})=\int_{-\infty}^{+\infty}d\lambda_{5}e^{i\lambda_{5}(t_{5}-\bar{t}_{5})}
\end{equation}
we obtain
\begin{equation}
 e^{-V\beta R \widetilde{V}_{eff}(t_{4},t_{5})}=\frac{1}{Z}\int d\lambda_{4}\int \lambda_{5}\int {\cal
 D}A_{\mu}^{\alpha}\;e^{-S[A_{\mu}^{\alpha}]+i\lambda_{4}(t_{4}-\bar{t}_{4})+i\lambda_{5}(t_{5}-\bar{t}_{5})}
\end{equation}
In order to compute the above path integral we use the saddle
point method. We split the fields into classical and quantum parts
\begin{equation}
A_{\mu}^{\alpha}=B_{\mu}^{\alpha}+Q_{\mu}^{\alpha}
\end{equation}
and
\begin{equation}
\lambda_{4,5}=b_{4,5}+q_{4,5}
\end{equation}
We expand the exponent in Eq. (47) up to quadratic terms in quantum
fields. The linear terms which are proportional to the equations of
motion, according to the saddle point method are required to vanish
(for details see Ref. \cite{two:5} ).
\begin{eqnarray}
&&\left(-\frac{\delta S}{\delta A_{4,5}^{\alpha}}+i
b_{4,5}\frac{\delta
\bar{t}_{4,5}}{\delta A_{4,5}^{\alpha}}\right)_{B_{4,5}^{\alpha}}=0 \\
&&-\left(\frac{\delta S}{\delta A_{1,2,3}^{\alpha}}\right)_{B_{\mu}^{\alpha}}=0\\
&&t_{4,5}-\bar{t}_{4,5}(B_{4,5}^{\alpha})=0
\end{eqnarray}
We assume that the background fields $B_{4,5}^{\alpha}$ are
constant and have different directions in the isospin space (note
that we have also assumed that $B_{1,2,3}^{\alpha}=0$ ). Of course
there is not a gauge transformation that can put the two gauge
field components in the same direction. However we can perform two
rotations in the isospin space. With the first we can put $B_{4}$
towards the generator $T_{3}$ and with the second we can put
$B_{5}$ on the plane defined by the generators $T_{3}$ and
$T_{1}$. Thus we can write
\begin{equation}
B_{4}=B_{4}^{3}T_{3}
\end{equation}
and
\begin{equation}
B_{5}=B_{5}^{3}T_{3}+B_{5}^{1}T_{1}
\end{equation}
We will look for a saddle point solution assuming that it is
possible to have $b_{4,5}=0$. From Eqs. (50) and (51) we see that
the classical field should satisfy the equations of motion for the
gauge fields:
\begin{equation}
-\left(\frac{\delta S}{\delta
A_{\mu}^{\alpha}}\right)_{B^{\alpha}_{\mu}}=\left(\partial_{\nu}F_{\mu\nu}^{\alpha}+g_{5}\epsilon^{\alpha
c b}F_{\mu\nu}^{b} A_{\nu}^{c}\right)_{B^{\alpha}_{\mu}}=0
\end{equation}
or
\begin{equation}
\epsilon^{\alpha c b}F_{\mu\nu}^{b} B_{\nu}^{c}=0
\end{equation}
Taking into account Eqs. (53) and (54) the only nonzero component
of the field tensor is
\begin{equation}
F_{45}^{2}=g_{5}B_{4}^{3}B_{5}^{1}
\end{equation}
From Eqs. (56) and (57), for $\mu=5$ and $\alpha=1$, we have
\begin{equation}
(B_{4}^{3})^{2}B_{5}^{1}=0
\end{equation}
and, for $\mu=4$ and $\alpha=3,1$, we have
\begin{equation}
(B_{5}^{1})^{2}B_{4}^{3}=0, \: \alpha=3
\end{equation}
and
\begin{equation}
B_{4}^{3}B_{5}^{1}B_{5}^{3}=0 , \: \alpha=1
\end{equation}
These three equations are satisfied only if $B_{4}^{3}=0$ or
$B_{5}^{1}=0$. However in general $B_{4}^{3}\neq 0$ (If we assume
that $B_{4}^{3}=0$ then we have only one gauge field component and
the case is trivial). So we must have $B_{5}^{1}=0$ and thus the
two gauge field components $B_{4}$ and $B_{5}$ have the same
direction in the group space.

Thus the saddle point we obtain has the form
\begin{equation}
B_{4}=B_{4}^{3}T_{3}, \: B_{5}=B_{5}^{3}T_{3}, \: b_{4,5}=0
\end{equation}

Now if we compute the path integral of Eq. (47), around this
saddle point, to one loop order, according to Ref. \cite{two:5},
we see that the result for the constraint effective potential is
in agreement with that of Eq. (24) for the one loop effective
potential.

\end{document}